\documentclass[journal]{IEEEtran}

\usepackage{epsfig}
\def\e{\begin{equation}}
\def\f{\end{equation}}
\def\=#1{\overline{\overline #1}}
\def\_#1{{\bf #1}}

\def\.{\cdot}
\def\x{\times}

\def\l#1{\label{eq:#1}}
\def\r#1{(\ref{eq:#1})}
\def\am{\left(\begin{array}{c}}
\def\amm{\left(\begin{array}{cc}}
\def\a{\end{array}\right)}

\def\E{\epsilon}
\def\M{\mu}

\def\.{\cdot}
\def\x{\times}

\begin{document}
%
\title{Patch antennas with new artificial magnetic layers}

\author{M.E.~Ermutlu$^{1,2}$,~\IEEEmembership{Member,~IEEE,} C.R.~Simovski$^3$,
        M.K.~K\"arkk\"ainen$^1$, P.~Ikonen$^1$,~\IEEEmembership{Student~Member,~IEEE},
        A.A.~Sochava$^4$,~\IEEEmembership{Member,~IEEE,}
        and~S.A.~Tretyakov$^1$,~\IEEEmembership{Senior~Member,~IEEE}
\thanks{\today \ version.}%
\thanks{$ ^1$Radio Laboratory / SMARAD, Helsinki University of Technology,
P.O. Box 3000, FI-02015 TKK Finland}
\thanks{$ ^2$Nokia Networks, P.O. Box 301, FI-00045, Finland}
\thanks{$ ^3$Dept. of Physics, St.~Petersburg Institute of
Fine Mechanics and Optics, Sablinskaya 14, 197101 St. Petersburg,
Russia}
\thanks{$ ^4$Radiophysics Department, St. Petersburg
Polytechnical University, Russia}}

\maketitle

\begin{abstract}
A new type of high-impedance surfaces (HIS) has been introduced by
C.R. Simovski et al. recently. In this paper, we propose to use such layers
as artificial magnetic materials in the design of patch antennas.
The new HIS is simulated and patch antennas partially filled by
these composite layers are measured in
order to test how much the antenna dimensions can be reduced. In order to
experimentally investigate
the frequency behavior of the material, different sizes of the
patches are designed and tested with the same material layer. Also the height of
the patch is changed in order to find the best possible position
for minimizing the antenna size. This composite layer of an
artificial magnetic material has made the antenna smaller while keeping the
bandwidth characteristics of the antenna about the same. About  $40\%$ of
size reduction has been achieved.

\end{abstract}

\begin{keywords}
artificial magnetic material, patch antenna, antenna
miniaturization, high-impedance surface.
\end{keywords}

\IEEEpeerreviewmaketitle

\section{Introduction}

\PARstart{H}{IGH} impedance surfaces for microwave frequency
antennas and other devices have been recently introduced and
actively investigated. These surfaces are mostly thought to be
applied for reducing surface waves, for controlling the
plane-wave reflection phase, and as artificial magnetic conductors
\cite{Siv1,Siv2}.

High-impedance surfaces are thin composite layers (as a rule
backed by a metal plane), whose
surface impedance has a parallel resonance at a certain
frequency or at several frequencies. Within the resonant band the
surface of the layer behaves as a magnetic wall
for normally incident plane waves. That is why the HIS are often
called {\em artificial magnetic conductors (AMC)}.
In practical cases, the thickness of an AMC layer at
resonance is much smaller than a quarter of the wavelength. It means that
the reflection coefficient of a plane wave from the structure can
be close to +1 at a very small distance from the ground plane.
This means that the electromagnetic interaction of a horizontal electric
current with an AMC can be
constructive even for very small distances from the current to the ground plane.
This is why AMC are prospective for designing low profile antennas
\cite{Siv1,motl,Yeo,homo}.

Infinite (practically very high) surface impedance means that the
tangential magnetic field component
is very weak at the surface of the wall. Tangential electric
field can be strong at this surface. The
reflection properties of a HIS depend on the frequency,
and the HIS resonant
band can be defined as a frequency range at which the phase
change of the reflected electric field is within certain limits
(usually from $-\pi /2$ to $+\pi /2$).
A HIS
conventionally possesses a rather narrow band of AMC
operation.

In general, these structures support surface waves, however there
is an important difference with the case of a simple metal-backed
dielectric layer: there are stop bands for surface waves.
Therefore, the same structures interacting with plane waves as HIS
operate as {\em EBG (electromagnetic band gap) structures} with
respect to the surface waves. Sometimes they are called as {\em 2D
PBG (photonic band gap) structures} which is not a proper term at
microwaves. Besides the high-frequency band gaps whose band
structure is related with the spatial periodicity of the AMC
layer, there can be one or two low-frequency ones, related with
the resonance of the surface impedance which can also hold in the
regime of the surface wave.

While the AMC are useful to design low-profile antennas \cite{Siv1,motl,Yeo,homo},
artificial material fillings are useful in the
design of small antennas, especially magnetic materials can be utilized
as a mean to reduce the antenna size \cite{Hansen,Mossad,Ziolk_APS}.

According to a simple theoretical model of a small patch antenna
as a resonator, the resonant frequency and impedance bandwidth
$BW$ of a small antenna depend on the relative permittivity and
permeability of the loading material as \e F_{r}
\sim\frac{1}{\sqrt{\varepsilon_{r}\mu_{r}}}, \quad BW
\sim\frac{\sqrt{\mu_{r}}}{\sqrt{\varepsilon_{r}}}. \l{sup}\f Here
$\sim$ means the proportionality. In practice, these expressions
are not very accurate, and the result of loading depends on the
antenna type, but still the tendency predicted by \r{sup} holds.

There has been some work on
the effects of  magneto-dielectric substrates on the bandwidth of
microstrip antennas \cite{Hansen,Mossad,Ziolk_APS}, but  these
results are contradictory. Papers \cite{Hansen} and \cite{Ziolk_APS}
conclude that the best filling is a magnetic material with large
$\mu$, although the authors of \cite{Mossad} conclude that the best filling
is a material with both $\epsilon\gg\epsilon_0$ and $\mu\gg\mu_0$.
Nevertheless, the usefulness  of magnetic filling is evident from
the known literature.

In paper \cite{Edvardsson} a material with a finite isotropic
permeability was considered as a loading material. Natural
magnetic materials at frequencies higher than  $100-500$ MHz are
ferrimagnetic crystals (for example, hexaferrites). These ferrites
are not very attractive for applications in small antennas since
they are lossy, heavy, and expensive. Probably because of this,
antennas with ferrite or ferromagnetic filling have been analyzed
mainly from the point of view of electrical control of their
resonant frequency and radiation pattern, see, e.g.,
\cite{Volakis}. Possible realizations of artificial magnetics at
$0.5-3$ GHz can be based on the resonant principle: Artificial
magnetic materials can be formed by small particles with a
resonant magnetic susceptibility. These should resonate at a
frequency which is close to the resonant frequency of an unloaded
antenna. Such artificial magnetic materials are practically known
in the frequency range $4-15$ GHz (the structures formed by the
so-called {\it split-ring resonators}). Such loadings have been
successfully tested with patch antennas in \cite{[4]}: A
$0.075\lambda$ size patch antenna with the bandwidth of $1.5\%$
has been reported in that paper.

In certain frequency regions,
high-impedance surfaces behave as layers of effective magnetic
materials, and can be used as magnetic fillings in the antenna
design.
AMC are not similar to natural isotropic
magnetics. These cannot be properly described in terms of
magnetic permeability. Formally, one can introduce an effective
permeability, but it will
strongly depend on the incident plane wave polarization and
the angle of incidence. This dependence is a spatial dispersion effect.
However, this effect does not forbid one to use
such composite layers as a magnetic filling material.
Recently, these artificial layers have been used to
reduce the size of antennas \cite{[2]} and filters \cite{[3]}.

The main idea of the present paper is to use an artificial
material which would possess the properties of both AMC and
magneto-dielectric composite \cite{[1]} within a rather wide
resonance band. In this paper, the new HIS \cite{[1]} playing also
the role of an artificial magnetic material (AMM) is tested under
a patch antenna. The choice of this material can be explained in
terms of the angular stability of the resonant frequency. The
known AMC introduced in papers \cite{Siv1,Siv2,Itoh} have the
resonance depending on the angle of incidence $\theta$ and wave
polarization (TE or TM). Therefore the interaction of such AMC
with currents on a real antenna cannot be completely constructive.
For example, let the working frequency of the patch antenna
correspond to the resonance of the conventional AMC illuminated by
a normally incident plane wave. Then the narrow part of the
angular spectrum of radiation centered at $\theta=0$ will interact
with the AMC as if it were a magnetic wall. For the other part of
the angular spectrum the conventional AMC is not a magnetic wall,
since its resonant frequency will be different
\cite{Tret,EL,PIER}. In the AMC used here this shortcoming is
absent as it was shown in \cite{[1]}. Notice, that in paper
\cite{Simov} another full-angle AMC has been suggested and
studied. This is a self-resonant grid on a simple metal-backed
dielectric layer. However, our choice of the artificial magnetic
material suggested in \cite{[1]} has additional advantages. First,
in this AMM there are vertical conductors (vias). We expect that
the TM-polarized surface waves are suppressed in this structure at
low frequencies due to the presence of vias in the same way as in
mushroom structures \cite{Siv1,Siv2}. Second, the resonance of the
structure  \cite{[1]} takes place at lower frequencies (compared
to the structure period) than the resonance of the AMC described
in \cite{Simov}. Both these advantages are important for patch
antennas. The second one is important for the antenna
miniaturization. The first one is crucial when one uses an array
of patches and the problem of mutual interaction appears.

The behavior of the AMM is investigated at different distances
from the patch antenna with different configurations. Also a
simple comparison method is introduced in order to find the
effective (averaged) permeability value of the layer. Results from
IE3D/FDTD simulations are shown and compared with the measurement
results.

\section{The new high impedance surface as an artificial magnetic material}

\begin{figure}
\centering \epsfig{file=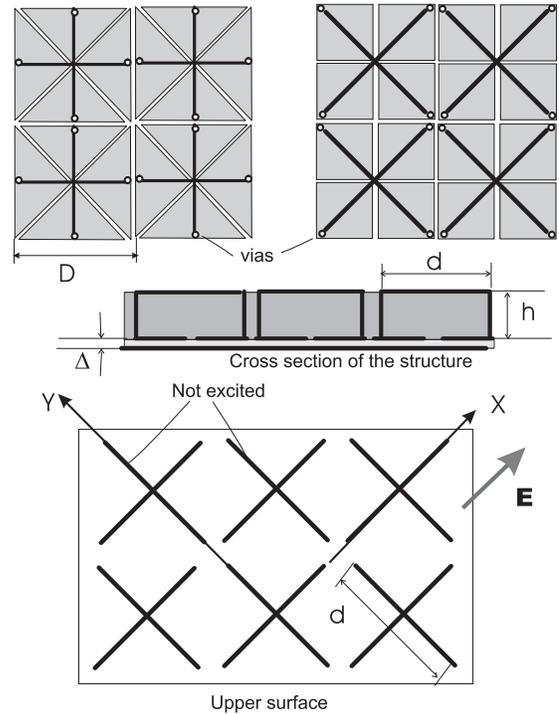, width=0.4\textwidth}
\caption{Elements of a HIS from \cite{[1]}. Top: two variants of
the structure geometry. Left: triangular patches. Right: square
patches. Middle: structure cross section along the $(x-z)$ or
$(y-z)$ loops. Bottom: upper surface of the structure containing
planar metal crosses. When $\_E=E\_x_0$ the $(y-z)$ loops are not
excited.} \label{cellnew}
\end{figure}

\begin{figure}
\centering \epsfig{file=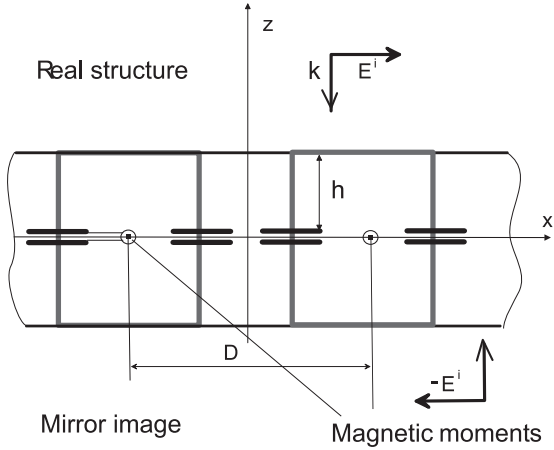, width=0.4\textwidth}
\caption{Equivalent scheme of a HIS.} \label{scheme}
\end{figure}

\PARstart{T}{HE}
structure introduced in \cite{[1]} is shown in
Fig.~\ref{cellnew}. This AMM can be considered as a 2D grid of
bulk unit cells with the horizontal period $D$. Every cell contains
two orthogonal loops of length $d$, so that the upper interface
represents an array of metal crosses on the surface of a dielectric
layer. Every
cross (whose ends are connected to metal vias) together with
vias and patches form two orthogonal loops (the loop length $d$ is
the same as the length of a cross side). So, the effective
vertical loop is formed by two vias perforating the dielectric
layer and a horizontal strip lying on the dielectric interface.
The loading capacitors are formed by metal patches and the ground
plane. The patch array is separated from the ground plane by a
thin dielectric layer.
The standard printed-board
circuit (PBC) can be used to prepare both patch array (which is
located on one side of a PBC) and the array of crosses (located on
the other one) if the thin layer has no metallization (teflon
film). The analysis of the structure impinged by a plane wave
becomes easier with the help of the image theory. Every real loop
complemented by its mirror image is a symmetrically loaded
rectangular loop with sizes $S=d \x P$, where $P=2(h+\Delta)$ (see
Fig.~\ref{scheme}). The electric field is zero at the loop
center (at the ground plane $z=0$). Therefore, the electric polarization
of the loop is negligible and it can be considered as a horizontal
magnetic dipole excited by an external magnetic field. Following
the image method, consider an array of loaded rectangular loops
illuminated by two plane waves from both sides of the array. Let
the plane wave be polarized so that $\_E=E\_x_0$ and the magnetic
field is directed along the $y$-axis. Then the loops in the $yz$
plane are not excited (see Fig.~\ref{cellnew}). The whole
structure behaves as an array of parallel loops lying in the
planes $xz$ within the dielectric layer excited by two plane waves
coming from $z=\infty$ and $z=-\infty$. Since there is no electric
dipole polarization of loops, the electric polarization of the
whole structure is practically that of the dielectric layer. The
reflected field is then the sum of the field produced by the
single dielectric layer of thickness $P$ (excited by two waves
impinging the layer from the top and the bottom) and the field produced by
the magnetic moments of loops. The magnetization of loops is
resonant due to the presence of capacitive loads, and the magnetic
response at the resonance is comparatively strong due to high
inductance of the loops. Loops are made from thin electric
conductors and their inductance is much higher than the effective
inductance of the conventional AMC (which is practically
determined by the thickness of the dielectric layer \cite{Siv2}).
This factor is responsible for rather low frequency of the
resonance and for a wider bandwidth (in parallel $LC$-circuits the
higher is the inductance the larger is the resonant band). An
analytical model of this structure and a comparison with the
results of numerical simulations (obtained with the HFSS package)
are presented in a recent paper by C. Simovski and A. Sochava
\cite{IEEE}.

\begin{figure}
\centering
\epsfig{file=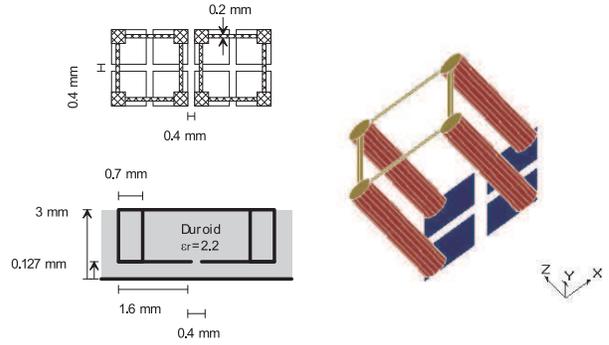, width=0.45\textwidth} \caption{Geometry and
dimensions of the manufactured AMM layer.} \label{fig1}
\end{figure}

The geometry of the
structure that we use in this paper
is different from the structure  proposed in \cite{[1]},
although the operational principle is the same.
This new structure is  shown in
Fig.~\ref{fig1}, where square horizontal loops replace the
conducting crosses of the structure \cite{[1]}. For simplicity,
let us consider the normal incidence of a wave whose electric
field is polarized along one side of the square horizontal loop.
Then the two vertical $C$-loaded loops (formed by two sides of the
horizontal loop, four vias connected to them and four capacitances
between patches and the ground plane) will be excited in every unit
cell of the structure. A horizontal loop as such is not excited by
the external magnetic field since this magnetic field is
tangential. The electric connection of the two vertical loops does
not change the operation and the theory developed in \cite{[1]}
basically remains valid.

To realize the material a two sided
TLY-5 film with the thickness of $0.127$ mm and the dielectric
constant $\E_r=2.2-j0.001$ has been used. This film separates  small
patches from the ground plane. Horizontal parts of the loops and
the patches are printed on the opposite sides of
a printed circuit board layer with the dielectric constant
$\E_r=2.2-j0.002$. Horizontal parts of the loops and the patches
are connected by via wires (round metal
cylinders)  as shown
in Fig.~\ref{fig1}. The whole manufactured structure contains $5\times 5$
unit cells and has the
dimensions of $20\times 20\times 3$ mm. This structure is a ``brick"
from which
larger samples of an AMM can be built.

\section{Antenna and the artificial magnetic material}

\begin{figure}
\centering \epsfig{file=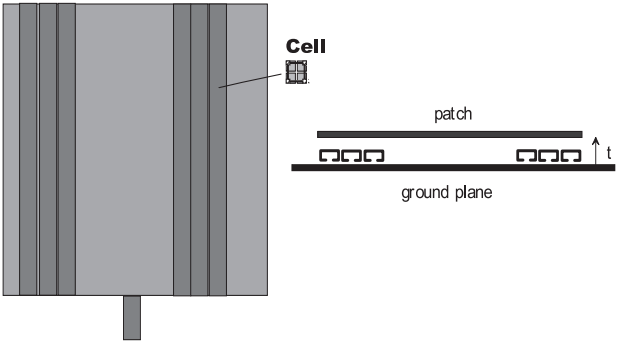, width=0.45\textwidth} \caption{A
patch antenna partially filled with an artificial material. Three
columns of the new AMM at each side of the antenna patch are
inserted. The first column is placed just under the side of the
patch where the currents are strong. The distance between the
patch and the ground plane is $t=4$, 5, and 6 mm. Different square
patches with the sizes $30\x30$, $40\x40$, $50\x50$, and $60\x60$
mm have been designed. The height from the ground plane $t$ and
the size of the patch are used as parameters to test the
artificial magnetic material.} \label{fig2}
\end{figure}

\PARstart{T}{O} test the performance of the new artificial
magnetic layers with patch antennas we design and study square-patch
antennas with different
sizes. For measurements, a large ground plane is used and for
simulations the ground plane is infinite. This is done in order to
exclude possible resonant effects of a finite ground plane. IE3D
software as well as an in-house developed FDTD code are used to simulate
antennas with and without the material filling. Also, for the sake of
comparison, we simulate the same antennas with an infinite dielectric
material layer inserted between the  ground plane and
the patch. The dielectric has the relative permittivity $\E_r =
2.2$ and the slab thickness is  $3$ mm (the
same as for the substrate used to manufacture the artificial
magnetic layer). The antenna is fed from a side using a
microstrip. The new AMM is introduced under the patch gradually in order
to save simulation time. In Fig.~\ref{fig2}, the configuration with
three columns of the
material under the patch antenna are shown. AMM blocks are located at the
two sides of the antenna symmetrically.

Antennas with the patch dimensions of $30\times 30$, $40\times 40$, $50\times 50$, and
$60\times 60$ mm have been considered. Since patches of different sizes
resonate at different frequencies from $2.5$ GHz to $4.5$
GHz, we could test the effectiveness of the material at different
frequencies. The height of the antenna changes
the field applied on the material which  also changes the
response of the  material layer.

\subsection{The effect of the patch height}

\begin{figure}
\centering \epsfig{file=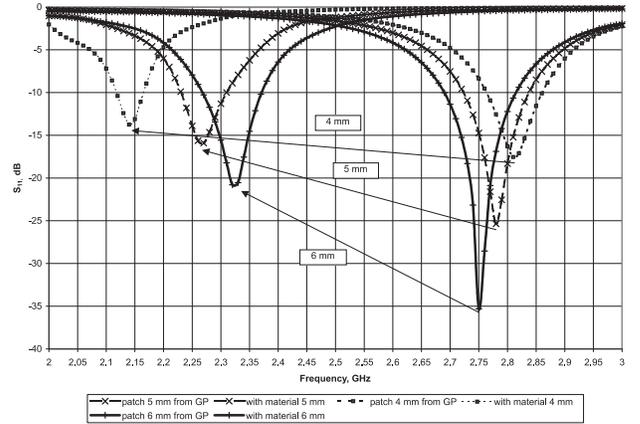, width=0.45\textwidth}
\caption{The return loss of the $50\times 50$ mm patch for
different antenna heights from $4$ to $6$ mm above the ground
plane. There is only one column of the material used.}
\label{fig3}
\end{figure}

The height of the patch antenna has been varied in order to find
the best reduction for the resonance frequency. In these
simulations, the $50\times 50$ mm antenna is chosen and only one
column of the material is placed under each side of the patch. The
return loss is compared for $4$, $5$, and $6$ mm heights of the
patch antenna (see Fig.~\ref{fig3}). The results show that the
effect of the material is enhanced when the patch is closer to the
material layer. Here we of course see also the effect of the
relative permittivity of the loading layer. If there would be no
magnetic behavior of the material, antenna would be thought as
loaded with a dielectric layer with $\E_r=2.2$ and the thickness 3
mm. In that case the resonant frequency drops from $2.82$ to
$2.28$ GHz for a  patch antenna with the patch at $4$ mm above the
ground plane. But with the magnetic material the resonant
frequency drops to $2.14$ GHz. Here, with only one column of the
artificial material we have a $24\%$ reduction in the resonant
frequency or, in other words, we have a patch antenna with a size
of $0.357\lambda$.

\subsection{The effect of the number of columns}

\begin{figure}
\centering
\epsfig{file=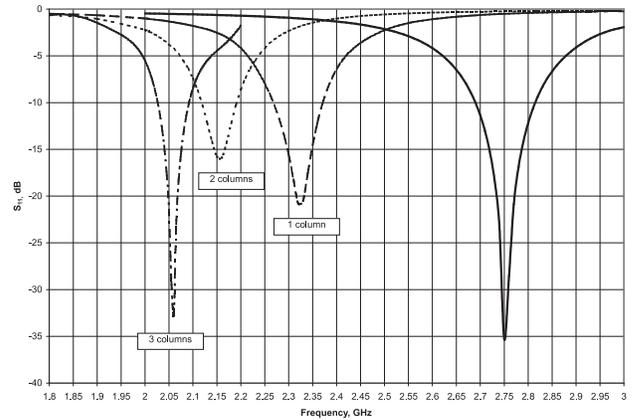, width=0.45\textwidth}
\caption{The effect of the new AMM on the return loss of the
antenna. The number of columns is increased to see the effect of the
increased volume fraction of the material filling.} \label{fig4}
\end{figure}

When an antenna is totally filled with such complex material, it becomes very
difficult to simulate it with IE3D, using the computer power we have.
This is because the AMC is made of a lot of small metal strips, patches, and vias.
Therefore we gradually fill the volume below the antenna. We start
to fill from the sides of the antenna where the currents
and magnetic fields are strong.
Here we
investigate the effect when the antenna is partially filled.
The number of columns is increased and
the effect in return loss is shown in Fig.~\ref{fig4} for the
$50\times 50$ mm patch. In this example the antenna patch is at $6$ mm above
the ground plane. The same exercise has been repeated with the patch
at $4$ mm above the ground plane. The results are listed in Table~\ref{table1}.

It is seen from the simulations that even with only one column we achieve
a 16\% reduction in  the antenna resonant frequency for the patch at $6$ mm
above the ground and $24\%$ for the $4$ mm height. With
three columns we achieve about 25\% and 33\% reduction for
heights  6 and 4 mm,  respectively.

\begin{table}
\renewcommand{\arraystretch}{1.3}
\caption{Resonant frequencies of the $50\times 50$ mm antenna with
and without AMM for the patch heights $4$ and $6$ mm.
The antenna resonant frequencies are also shown when there is
an infinite dielectric material slab under the antenna patch with the $3$ mm thickness  and
$\E_r=2.2$.}
\label{table1} \centering
\begin{tabular}{|c||c||c|}
\hline $t=4$ mm & Res. freq. (GHz) & Reduction \% \\ \hline No
material & 2.82 & 0.00 \\ \hline No AMM $\E_r=2.2$ & 2.28 & 19.15
\\ \hline $1$ column & 2.14 & 24.11 \\ \hline
$2$ columns & 2.00 & 20.08 \\ \hline $3$ columns & 1.89 & 32.98
\\ \hline \hline
$t=6$ mm & Res. freq. (GHz) & Reduction \% \\ \hline
 No material & 2.75 & 0.00 \\ \hline No AMM $\E_r=2.2$ & 2.45 & 10.91
\\ \hline $1$ column & 2.32 & 15.64 \\ \hline $2$ columns & 2.16 &
21.45 \\ \hline $3$ columns & 2.06 & 25.09 \\
\hline
\end{tabular}
\end{table}

\subsection{The effect of the patch size}

\begin{figure}
\centering
\epsfig{file=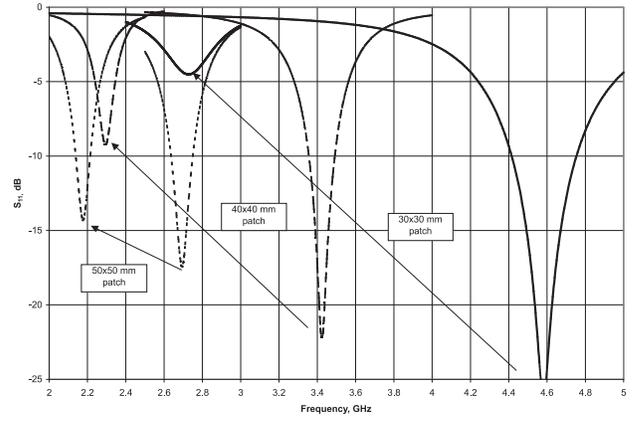, width=0.45\textwidth}
\caption{The effect of the AMM on the resonant frequencies of $30\times
30$, $40\times 40$, $50\times 50$, and $60\times 60$ mm patch antennas
is shown. There is only one column of AMM used. Antenna patches  are all
at $4$ mm above the ground plane.} \label{fig5}
\end{figure}

In order to see the effect of the new HIS at different
frequencies, different-size antennas are simulated with and
without the material. Resonant frequencies of antennas with an
infinite dielectric material layer placed under the patch antenna
with the $3$ mm height and  $\E_r=2.2$ are also shown for
comparison. There is a 1 mm distance between the material layer
and the antenna ground plane. First, only one column is placed
under $30\times 30$, $40\times 40$, $50\times 50$, and $60\times
60$ mm patch antennas with the height of 4 mm, then for $30\times
30$ mm, $40\times 40$, and $50\times 50$ mm patch antennas three
columns are placed. In Fig.~\ref{fig5}, return loss of patch
antenna is shown with and without materials. Also in
Table~\ref{table2} we show the reduction in the resonant
frequencies  with a dielectric material and when the AMM is
inserted. The results clearly show that the magnetic properties of
this material sample are stronger at higher frequencies (near 4.5
GHz) than at 3 and 2 GHz. On the other hand, we see that the
resonance is quite broadband, as the effect is rather strong even
far from the resonance of the AMM.

\begin{table}
\renewcommand{\arraystretch}{1.3}
\caption{Resonant frequencies of different-sized patches and
reduction in the resonant frequencies compared to a patch antenna where
there is only a dielectric material with  $\epsilon_r=2.2$, one column of
the artificial material,
and three columns of the material at both sides of the patch.}
\label{table2} \centering
\begin{tabular}{|c||c||c|}
\hline Patch size (mm) & Reduction \% $\E_r=2.2$ & Reduction \% 3
columns \\ \hline
50 & 10 & 19 \\
\hline 40 & 17 & 32 \\ \hline 30 & 17 &  40 \\
\hline
\end{tabular}
\end{table}

\subsection{Current distribution and the radiation pattern}

The current distribution and the radiation pattern of the antenna are
shown for the $30\times 30$ mm patch antenna with the patch at $4$ mm above the
ground plane. The material has been placed in three columns. The
antenna resonates at $2.7$ GHz. It is clearly seen that the current
distribution on the patch has an effect on the material and the current is
rotating on the loop part of the material. The calculated radiation pattern is
shown in Fig.~\ref{fig7}. The gain of the antenna is $5.3$ dBi.  The shape of the
pattern is similar to that of a usual patch antenna, so we can conclude that
the AMM material is working as expected.

\begin{figure}
\centering \epsfig{file=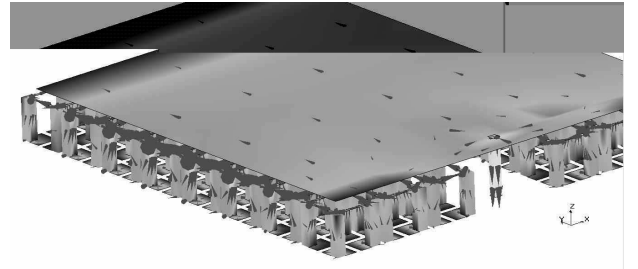, width=0.45\textwidth}
\caption{Vector current distribution of the $30\times 30$ mm patch
antenna with three columns of AMM. The scale is going from red
(maximum) to dark blue (minimum, $-40$ dB).} \label{fig6}
\end{figure}

\begin{figure}
\centering \epsfig{file=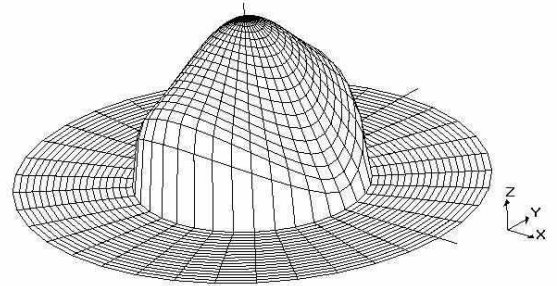, width=0.45\textwidth}
\caption{Radiation pattern of the $30\times 30$ mm antenna with
three columns of AMM at $2.7$ GHz. The maximum gain is 5.28 dBi.}
\label{fig7}
\end{figure}

\section{Estimation of the equivalent effective permeability}

As has been already noted, the artificial
magnetic layers built as HIS surfaces cannot be properly described in terms of
magnetic permeability, since that would
strongly depend on the incident plane wave polarization and
on the angle of incidence. However, it is possible to
introduce an equivalent effective averaged permeability
for this particular application, comparing the performance of
the actual antenna with an artificial layer and calculated
results for the same antenna filled by a uniform and
isotropic magnetic material.

This has been done using IE3D simulations. First, we simulate
antenna with different sizes of the patch
filled by an infinite slab
of an isotropic magneto-dielectric material (thickness 3 mm). The
material parameters are changed from $\E_r=2.2$,  $\M_r=1$ to
$\E_r=2.2$, $\M_r=4$. Then the same antennas
are filled by the AMM are simulated. Resonant frequencies of these results
are recorded and graphs are drawn for the calculated
frequency shifts. Comparing the resonant
frequencies, the equivalent permeabilities are identified.
These equivalent permeability values are underestimating the
actual permeability values of the AMM since AMM is not
uniformly filling the volume and it is not isotropic.
The field applied to the AMM is not uniform,
therefore different parts of the AMM sample are excited
differently. But this comparison gives a
clear and easy way of understanding the effective permeability of
the material and could be used as a helping design tool in the future.

\begin{figure}
\centering
\epsfig{file=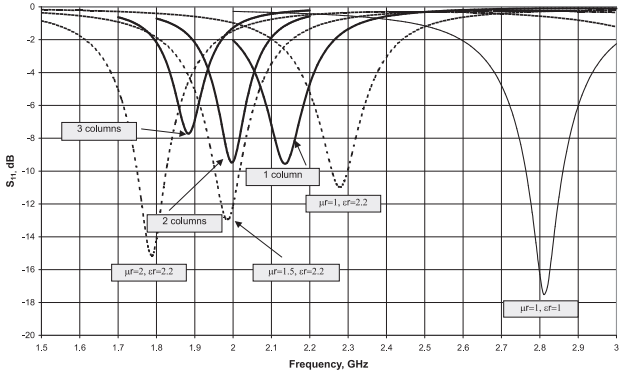, width=0.45\textwidth}
\caption{Comparison of the response of antennas filled by the
artificial magnetic material and by uniform isotropic magneto-dielectrics. } \label{fig8}
\end{figure}

In Fig.~\ref{fig8} one can see the return loss of the $50\times 50$
mm patch at $4$ mm above the ground plane with a 3 mm material layer with different $\M_r$
values. In that figure, the results for the
same antenna with three columns of AMM are also shown. For example,
for three columns it is clearly seen
that the equivalent averaged relative permeability value is between $1.5$ and $2$. With this
method the effective material permeability value is estimated as  $\M_r
=1.7$ at the resonant frequency of the $50\times 50$ mm patch
and 2.43 for the resonant frequency of the $30\times 30$
mm patch.

\begin{figure}
\centering
\epsfig{file=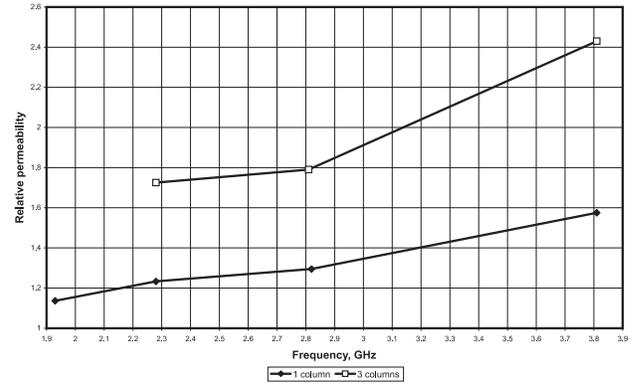, width=0.45\textwidth}
\caption{Calculated effective relative permeability values of the AMM for one
and three columns as functions of the frequency.} \label{fig9}
\end{figure}

In Fig.~\ref{fig9}, the averaged effective material permeability is shown
as a function of the frequency. The frequency values are the values when the
infinite material layer has  $\E_r=2.2$,  $\M_r=1$. When the AMM is
introduced, the resonant frequency drops. From this difference the
effective relative permeability values are calculated. This procedure is done
for one and three columns of the material and the results are shown also in Table~\ref{table3}.

\begin{table}
\caption{Calculated effective relative permeability values of the AMM for one
and three columns as functions of the frequency.} \label{table3}
\centering
\begin{tabular}{|c||c||c|}
\hline Patch size (mm) & Calc. $\M_r$, one column & Calc.
$\M_r$, three columns \\ \hline
60 & 1.136 & - \\
\hline 50 & 1.233 & 1.725 \\ \hline 40 & 1.295 & 1.79 \\ \hline 30 & 1.575 & 2.43 \\
\hline
\end{tabular}
\end{table}

\section{Comparison between measurements and simulations using  FDTD and IE3D}

\begin{figure}
\centering
\epsfig{file=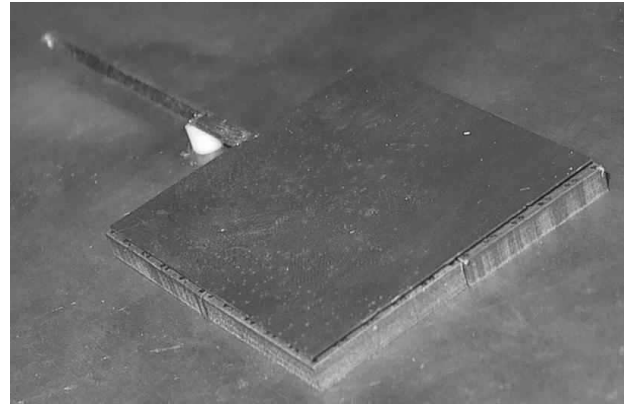, width=0.45\textwidth}
\caption{Photo of the measured antenna filled with an artificial
magnetic layer.} \label{fig10}
\end{figure}

An antenna with the  patch size $40\times 40$ mm has been built.
The patch  is first positioned at 6 mm above the ground plane,
then at 4 mm above the ground plane. It is filled with the
artificial material layer of the dimensions $40\times 40\times 3$
mm (Fig.~\ref{fig10}). The antenna in Fig.~\ref{fig10} has also
been simulated with a 3D FDTD code. In FDTD, the particles are
constructed from joint thin wires and small plates of
approximately same size as in measurements (within limits of
finite cell sizes).

\begin{figure}
\centering
\epsfig{file=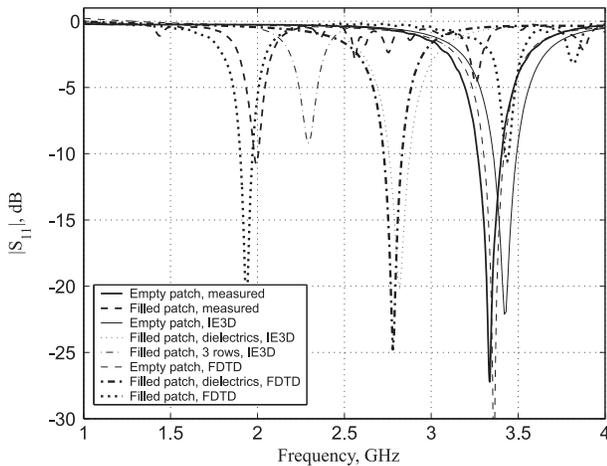, width=0.45\textwidth}
\caption{Comparison of simulated  and measured results.} \label{fig12}
\end{figure}

As can be seen from Fig.~\ref{fig12}, the results are similar even
though there are differences in the resonant frequencies. Of
course in simulations we have an infinite ground plane, and only
three columns of the artificial material partially fill the
antenna volume in the case of IE3D simulations. On the other hand,
in measurements we have had a large but finite ground plane, and
the material sample is of the same transverse dimensions as the
patch.

\section{Conclusions}

A new artificial magnetic material layer has been tested in order to shrink patch antenna
dimensions. In
simulations a $0.34 \lambda$   antenna and in practice a $0.38 \lambda$   antenna
have been realized, with the 6-dB bandwidth of 4.35\%.
The antenna bandwidth of these reduced-size antennas
is practically of the same order as for usual air-filled patch antennas with the
patch size $0.5 \lambda$, that resonate at the same frequency.
Thus, we have demonstrated in practice a technique to  miniaturize patch antennas
with the use of a certain
high-impedance surface (working as an artificial magnetic material layer) without worsening the
bandwidth.


\end{document}